\documentclass[useAMS,usenatbib]{mn2e}

\title[Strong magnetic field in W75N OH maser flare] {Strong magnetic field in W75N OH maser flare}
\author[V. I. Slysh and V.~Migenes]
{V.I.~Slysh $^1$\thanks{E-mail: vslysh@asc.rssi.ru} and V.~Migenes $^2$\\
$^1$Astro Space Centre, Lebedev Physical Institute, Profsoyuznaya str. 84/32, Moscow 117997, Russia\\
$^2$Department of Astronomy, University of Guanajuato, Apdo Postal 144, CP36000 Guanajuato, Mexico}

\date{Accepted 2006 February 15.
      Received 2005 December 30;
      in original form 2005 December 20}

\pagerange{\pageref{firstpage}--\pageref{lastpage}}
\pubyear{2005}

\begin{document}

\maketitle

\label{firstpage}

\begin{abstract}
A flare of OH maser emission was discovered in W75N in 2000. Its location was determined with the VLBA to be within 110~au from one of the ultra-compact
H~II regions, VLA2. The flare consisted of several maser spots. Four of the spots were found to form Zeeman pairs, all of them with a magnetic
field strength of about~40 mG. This is the highest ever magnetic field strength found in OH masers , an order of magnitude higher than in typical OH masers. Three possible
sources for the enhanced magnetic field are discussed: (a) the magnetic field of the exciting star dragged out by the stellar wind; (b) the general interstellar
field in the gas compressed by the MHD shock; and (c) the magnetic field of planets which orbit the exciting star and produce maser emission in gaseous envelopes.
\end{abstract}


\section{Introduction}
The formation of stars is proceeding with the outflow of gas and the formation of disks. The disks serve as reservoirs of matter for accretion onto the protostar and
increasing its mass. The remnants of the disk may be the home of protoplanetary systems. In this process magnetic fields play a crucial
role determining disk rotation and the rotation of the protoplanets, and the rate of accretion (Uchida \& Shibata\ 1985,  Shu et al.\ 1994). OH masers
 are very sensitive indicators of the magnetic
fields due to the large Lande {\it g\/}-factor compared to other molecules detectable in protostellar objects. Previous observations have determined magnetic
field strengths in OH-maser spots in the range from 1 to 8 milligauss (mG) (Fish et al.\ 2005). One of the best studied
star-forming regions is  W75N which is embedded in a dense molecular cloud (Hunter et al.\ 1994). In the core of the cloud there are several
young massive stars exciting
compact H~II regions. These are related to OH, H$_2$O and CH$_3$OH masers (Haschick et al.\ 1981, Hunter et al.\ 1994,
Minier, Conway \& Booth\ 2001). High angular resolution mapping of the masers has shown that the
 maser emission is concentrated in clusters which tend to be associated with the known ultra-compact H~II regions VLA1 and VLA2 (nomenclature is from
 Torrelles et al.\ (1997)). Most of the OH
 maser spots are associated with VLA1, forming an elongated arc with a velocity gradient, which can be modelled by a rotating disk (Slysh et al.\ 2002).
 The magnetic field in these spots has a typical OH-maser value of several mG. Near VLA2 only two 1665-MHz OH maser spots were found in 1998.
 In 2000 a strong flare of OH maser emission from W75N was discovered. It was detected in VLBA and EVN observations, as well as in our single dish
 observations in April 2001, at the declining phase of the flare. It is possible that this flare was a precursor of an even stronger, 1000-Jy flare, which
  started two years later (Alakoz et al.\ 2005). The OH maser W75N is  unusual in
 showing several polarized spectral features with a high degree of linear polarization (Slysh et al.\ 2002). Here we report on high angular resolution
 VLBI observations of the OH maser in W75N, in which several maser spots with a very strong magnetic field, were found to accompany a major flare of OH
 maser emission.

\section{Observations}

     The new observations of the OH maser W75N were conducted on 2001 January 01 with the VLBA in the snapshot mode of 6-min duration. The velocity resolution
 was 0.176 km s$^{-1}$, with 256 spectral channels covering 45 km s$^{-1}$ in each of the OH main lines at 1665 and 1667 MHz. In addition, we reduced
 and analyzed the EVN
 observations, with the same velocity resolution, from the EVN archive (project EP037b)
 of 2000 September 27, performed three months earlier, with a velocity coverage of 90 km s$^{-1}$. Other relevant observations
 are available from the public VLBA archive. These are for 2000 November 22 and 2001 January 6, and were recently published
 (project BF064, Fish et al.\ 2005). These observations are especially interesting because they were conducted two months before and 5 days after our
 observations. The VLBA archive observations had the same spectral resolution as ours but the velocity coverage was only 22.5 km s$^{-1}$ which is
 a factor of two less. This velocity coverage was not sufficient for detecting widely separated Zeeman pairs with strong magnetic field.
 However at 1667 MHz where the {\it g\/}-factor is 0.6 of 1665-MHz {\it g\/}-factor the velocity coverge is large enough even for the
 Zeeman pairs with the strong magnetic field which are the topic of the present study. All three sets of data have been obtained during the maximum
 phase of the OH maser flare. The data were reduced in the standard way using NRAO software package AIPS. Images of W75N were constructed for all spectral
 channels which had enough signal. Only those maser spots which were present in at least two spectral channels were considered as detected. Gaussian
 fitting and beam deconvolution were carried out using the task SAD of AIPS. Most of the maser spots were unresolved by the synthesized beam. The absolute
 position given in Table 1 was measured through fringe rates using AIPS task FRMAP.

\section{Results}

\subsection{Spectra and variability}
 Both the EVN spectrum of 2000 September 27 and the VLBA spectrum of 2001 January 1, taken 96 days later, of W75N show all the spectral features which were
 identified in the 1998 spectrum by Slysh et al.\ (2002). Fig.~1 shows the 1665-MHz spectra in Stokes {\it I\/}. The labels are the same as in the 1998 spectrum.
 Additionally, two new
 strong spectral features have appeared since the 1998 observations, at the low-velocity side of the spectrum. They are {\it P1\/} ({\it P}recursor),
 with the flux density of 120~Jy becoming the strongest feature, and {\it P2\/}. On the other
 hand, two relatively strong spectral features from the 1998 spectrum, {\it J\/} and {\it K\/}, became a factor of 3 weaker  in 2000-2001 spectra. In fact the
 feature {\it K\/} is so weak compared to {\it P2\/} which is at nearly the same velocity, that it is not seen in the spectrum; however it has been
 found on maps as is shown in Fig.~2. It is also
 evident from the spectra of Fig.~1 that the new 'flare' features were rapidly evolving in about a three months time interval, between 2000 September 27 (EVN)
 and 2001 January 1 (VLBA): {\it P1\/} has increased by a factor of 2, and {\it P2\/} has strengthened even more, almost by an order of magnitude. Four
 months later, on 2001 April 12 {\it P1\/} and {\it P2\/} have become weaker by a factor of 5 and 2 respectively as observed with the Bear Lake
 64-m single dish telescope (Alakoz et al.\ 2005). In the same time interval the rest of the spectral features, from A to H, remained unchanged. All constant
 features are connected t0 the ultra-compact H~II region VLA1 while the variable features are connected to VLA2 (see next subsection).

\begin{figure}
\hspace{-5mm}
\includegraphics{spectra00,01.eps}
\vspace{70mm}
\caption{EVN (27sep2000 full line) and VLBA (1jan2001 dashed line) spectra showing new features {\it P1\/} and {\it P2\/}.
The labelling of the features from {\it A\/} to {\it J\/} is from 1998 VLBA spectrum in Slysh et al.\ (2002). The vertical axis is the correlated
flux density on the short base lines Effelsberg-Torun and Los Alamos-Pie Town for the EVN and the VLBA instruments, respectively.}
\end{figure}

\subsection{Maps}

\begin{figure}
\vspace{0mm}
\hspace{-10mm}
\includegraphics{mapvla2.eps}
\vspace{80mm}
\caption{A map of OH maser spots near the ultra-compact H~II region VLA2. The 'plus' signs show the maser spots. $J$ and $K$ are the only spots which
were present near VLA2
in the 1998 map. $P1$, $P2$, and $P3$ represent the new 'flare' features, and the crosses indicate the positions of Zeeman pairs. The large circle
shows position of the ultra-compact H~II region VLA2; the size of
the circle does not correspond to the size of the VLA2 which is unresolved with 0.1{\arcsec\/} beam in Torrelles et al.\ (1997).}
\end{figure}

 The radial velocities of {\it P1\/} and {\it P2\/} are only about 0.7 km s$^{-1}$ lower than those of {\it J\/} and {\it K\/}, respectively.
 One possibility is that the flare occurred at the site of {\it J\/} and {\it K\/}. Mapping results however do show that {\it P1\/}, {\it P2\/}
 and {\it J\/}, {\it K\/} are separate   features, and all of them are present in the 2001 image (Fig.~2). Both {\it J\/} and {\it K\/}
 remain at the same position as in 1998, but are much weaker. {\it P1\/}, {\it P2\/} and a weaker feature {\it P3\/} have emerged not far from {\it J\/}
 and {\it K\/}, closer to the continuum source VLA2. Compared to the 1998 map (Slysh et al.\ 2002) several additional spots have been detected,
 partly due to a higher sensitivity of new observations.  One of the new features has completed a Zeeman pair with the spot {\it H\/} (see next subsection).
The other features are really new because they are related with the flare which took place between 1998 and 2000. Also, more accurate absolute positions
of OH spots were obtained and are given in Table~1, as well as the positions of the 1667 MHz spots relative to the 1665 MHz spots. The new map
presented in Fig.~2 shows the position of the
OH maser spots relative to the position of continuum source VLA2 (adopted from Shepherd et al.\ (2004)). The combined error of the absolute positions of the
continuum source VLA2 and OH masers is estimated here as 40 milliarcseconds (mas) while the separation between VLA2 and the nearest OH maser spots is 55~mas.
The relative position errors of the maser spots are typically less than 1~mas.

\subsection{Zeeman splitting}

The seven Zeeman pairs found in W75N 2000-2001 spectra, are given in Table 1. All of them have been identified as pairs of Zeeman components based on positional
coincidence to within  a fraction of the beam. The pairs are $\sigma$-components, with opposite sense of circular polarization. The requirement of the
positional coincidence between Zeeman components is very stringent and is based on the nature of Zeeman splitting. Each molecule is emitting all Zeeman
components simultaneously as the molecule is precessing in the magnetic field. The precession causes modulation of the emitted wave which results in
splitting of the monochromatic spectrum into several components with different amplitude, frequency, and polarization. Thus each group of molecules with
the same orientation relative to the magnetic field direction emits the same pattern of Zeeman spectrum. In the case of OH molecules this means that
every molecule emits both RCP and LCP $\sigma$-components, at their respectively shifted frequencies. On the map positions of true $\sigma$-components
must coincide. Since in W75N at 1665 MHz most of the maser spots are unresolved by the VLBA beam and appear as point-like sources, any observed position
difference between $\sigma$-components can be attributed to position measurement errors caused by low signal-to-noise ratio or to a misalignment between
RCP and LCP beams. In this study we adopt 5 mas as the maximum separation between Zeeman pairs which is small enough to exclude chance coincidences, even
in the presence of clustering of the maser spots. Fig.~3 shows one of the new Zeeman pairs $Z4$ in the 1665 MHz
spectrum of W75N, composed of the LCP-spot  {\it P3\/} at --0.48 km s$^{-1}$
(grey scale) and the RCP-spot at 24.40 km s$^{-1}$ (contours). Both spots are unresolved by the VLBA beam, and their positions coincide within 1.6 mas
which is comparable to the position measurement errors. This pair of the maser spots can be regarded as a pair of true Zeeman $\sigma$-components as well
as the rest of the pairs in Table 1.
\begin{table*}
\centering
\begin{minipage}{140mm}
\caption{OH Zeeman pairs in W75N}
\begin{tabular}{@{}crrrrrcccc@{}}
\hline
\multicolumn{1}{c}{Zeeman}&\multicolumn{3}{c}{RCP}&\multicolumn{3}{c}{LCP}\\
pair&$\Delta{RA}$&$\Delta{Dec}$&Velocity&$\Delta{RA}$&$\Delta{Dec}$&Velocity&B&Separation&Associated\\
number&(mas)&(mas)&(km s$^{-1}$)&(mas)&(mas)&(km s$^{-1}$)&(mG)&(mas)& H~II region\\
\hline
$Z1$%
\footnote{$Z1$ is spot {\it A\/}; the absolute position of {\it A\/} was measured to be: $RA=20^{h}38^{m}36.416^{s};
  ~Dec=42{^\circ}37{\arcmin\/}34.42{\arcsec\/} (\pm{0.01}\/{\arcsec\/})$ (J2000)}&0.0&0.0&12.45&0.16&0.03&9.28&5.4&0.2&VLA1\\
$Z2$&205.42&666.92&8.07&206.91&668.90&5.43&7.5%
\footnote{1667 MHz}&2.5&VLA1\\
$Z3$&488.80&1345.93&7.24&489.13&1345.29&5.13&3.6&0.7&VLA1\\
$Z4$&765.77&-178.12&24.40&766.15&-177.75&-0.48&42.2&1.6&VLA2\\
$Z5$&768.89&-135.24&19.80&767.59&-132.00&4.76&42.5%
\footnote{1667 MHz (Fish et al.\ 2005, table 15). This Zeeman pair was probably overlooked by the authors, or dismissed as showing too large velocity
separation.}&3.5&VLA2\\
$Z6$&770.30&-122.33&29.27&770.89&-125.00&4.71&41.6%
\footnote{from EVN data}&2.7&VLA2\\
$Z7$&770.41&-129.98&21.17&772.36&-128.62&8.42&36.3&2.4&VLA2\\
\hline
\end{tabular}
\end{minipage}
\end{table*}

\begin{figure}
\hspace {-17mm}
\includegraphics{Z4.eps}
\vspace {90mm}
\caption{Zeeman pair $Z4$ with a magnetic field strength of 42.2 mG. RCP(contours) at 24.40 km s$^{-1}$ and LCP(grey scale) at $-0.48$ km s$^{-1}$. Grey scale
maximum is 2 Jy beam$^{-1}$, contour peak at 0.4 Jy beam$^{-1}$, and the level intervals are 0.04 Jy beam$^{-1}$. The beam is 12.84 mas$\times$4.29 mas at
PA=$-28{^\circ}$.}
\vspace {0mm}
\end{figure}

In addition to Zeeman pairs found in 1998 observations (Slysh et al.\ 2002) several new ones are reported here. The first is the Zeeman counterpart
to spot H, at 7.24 km s$^{-1}$ corresponding to a magnetic field strength of 3.6 mG ($Z3$ in Table~1). This pair is associated with the ultra-compact H~II region
VLA1 as well as two other pairs $Z1$ and $Z3$, in Table~1. Four new Zeeman pairs are found in association with the ultra-compact
H~II region VLA2. The pairs $Z4-Z7$ associated with VLA2 are quite remarkable. They show a very strong magnetic field, from 36.3 to 42.5~mG. Another
property of the VLA2 Zeeman pairs is the apparent uniformity of the magnetic field across at least 100~au (50~mas): the field strength is equal
within 10 per cent in four spots, and has the same direction. Such a large field is unusual for OH masers, typically it does not exceed 8~mG
(Fish et al.\ 2005). Two outstanding Zeeman pairs in the survey of Fish et al.\ (2005) in W51e2 have magnetic field strength of 20~mG  which is only
half of the field of the new Zeeman pairs in W75N.

\subsection{Past observations of flares with strong magnetic field}

There have been several reports of flares for the high-velocity 1667-MHz masers.  For example, spectrum of W75N obtained in 1973 at Onsala shows so called 'Elld\'{e}r
transient' in the right circular polarization in the velocity range from 20 to 30 km s$^{-1}$ (Yngvesson et al.\ 1975). The 'high-velocity'  instead
may be a result of the large Zeeman velocity shift in the strong magnetic field, such as reported in this paper.
Hutawarakorn, Cohen \& Brebner\ (2002) located a flare in 1986 near VLA2 with the lower resolution of MERLIN. In addition to Zeeman pairs $Z_4$ and $Z_3$
listed in their table~7 there is probably one more Zeeman pair in table~4
consisting of an RCP-component at 22.5 km s$^{-1}$ and an LCP-component(s) at radial velocities 1.2, 2.6, or 4.2 km s$^{-1}$, with a separation
between 8 and 21~mas which is less than the estimated relative position error of 45~mas. If real, this 'flaring' Zeeman pair has a magnetic field strength of about
55~mG. The location of the 'flare' is within 15 mas from our 1667-MHz Zeeman pair $Z6$ in Table~1. These authors made an interesting comment that a
powerful magnetic
field might be an explanation, among others, of the high velocities in the flare. The 'flare' was not present in our 1998 data.

\section{Discussion}

\subsection{Relation to water masers}

Similar to OH-masers, the water masers in W75N are located in two clusters around VLA1 and VLA2. Torrelles et al. (2003) have found a shell of water
masers around the ultra-compact H~II region VLA2 with a radius of 160~au. The shell is expanding with a velocity of 28 km s$^{-1}$, perhaps episodically,
as one in a recurrent outflow. The high magnetic field OH maser spots $Z4-Z7$ are located very close to VLA2 (Fig.~2), at a distance of
55 $\pm{40}$~mas,
or at the projected distance of 110 $\pm{80}$~au. Therefore, the OH masers may well be located in the same shell as the water masers. The magnetic field
in water masers associated with star-forming regions is typically around 100~mG (Sarma et al.\ 2002). Somewhat higher magnetic field, up to 500~mG, was
measured in Cepheus~A water maser (Vlemmings et al.\ 2005). For W75N Fiebig \& G\"{u}sten\ (1989) give an upper limit of the line-of-sight magnetic field
strength of
34~mG for the maser line at 12.3 km s$^{-1}$. With a single-dish telescope they could not relate this line with a particular ultra-compact H~II region.
Both VLA1 and VLA2 might have a line with such radial velocity (see Torrelles et al.\ 1997). Assuming that 100~mG is a typical value for the magnetic
field strength of water masers, which is an order of magnitude higher than in typical OH masers, we find that the field
is of the same order as in the OH maser flare of W75N reported here.

\subsection{Shock origin of the maser flare}

The ultra-compact H~II regions VLA1 and VLA2 are excited by massive B-stars of about 10 M$_{\sun}$ (Shepherd et al.\ 2004).  Apparently, the exciting star
for VLA2 is  more active than that of VLA1. The appearance of new, strong maser features {\it P1\/} and {\it P2\/} near VLA2, as well as the
almost simultaneous dimming of nearby features {\it J\/} and {\it K\/} was interpreted as a passage of an MHD shock from {\it J\/} and {\it K\/}
to {\it P1\/} and {\it P2\/} (Alakoz et al.\ 2005). The shock was probably generated by the exciting star of VLA2, and was propagating in the gas of the
stellar wind. Later, the shock reached another site of OH molecule concentration and produced an even more powerful flare of the maser emission, with a
flux density of 1000~Jy (fig. 5 in Alakoz et al.\ (2005)). The results of observations of this flare will be a subject of a separate publication.

\subsection{The source of the magnetic field}

\subsubsection{Star}

The magnetic field strength measured in OH maser features may have its origin in the 10 M$_{\sun}$ exciting stars of VLA1 and VLA2. The flare features with a
magnetic field strength of 40~mG are located at the projected distance of 110~au from VLA2. The star may be emitting a stellar wind which has $r^{-2}$
distance density dependence. If the magnetic field energy density $B^{2}/8{\pi}$ has a similar dependence on distance as would be the case of the energy
equipartition then the magnetic field $B$ would scale with distance as $r^{-1}$. Hence, 40~mG at 110~au from the star would correspond to 110~G at the base
of the stellar wind, at a distance 6$\times{10^{11}}$ cm from the centre of the star, close to the star surface. If the true distance is larger
than the projected distance the surface magnetic field of the star would be
accordingly stronger, say 500~G. Such a field seems to be quite
reasonable for young massive stars. Recently, a magnetic field of 1300~G was measured in one of the Orion Trapezium cluster O-stars $\vartheta^1$ Orionis C
(Wade et al.\ 2006).
 Other (non-flare) Zeeman pairs which show  weaker magnetic field strengths, are located farther away from the stars, at the
projected distance of about 2000~au. At this distance the stellar wind's magnetic field strength is a factor of 20 weaker, that is about 2.5~mG. This is a typical
value for the magnetic field strength in the non-flare OH maser features. Therefore, the high value of the magnetic field strength in the flare features is due to their
proximity to the exciting star.

\subsubsection{Dense gas}

Another possible source for the strong magnetic field in OH masers is a shock compression of gas. The Zeeman pairs are located in the common shell
with water masers. If we assume the magnetic field strength in water masers to be 100~mG and a gas density of $10^{9}$ cm$^{-3}$ then for OH masers with a
magnetic field strength of 40 mG the gas density must be 1.6$\times{10^{8}}$ cm$^{-3}$, if the density scales as  {\it $B^2$}. This is a factor 50 to 100 higher
than theoretical density estimates for OH masers, although Gray, Hutawarakorn \& Cohen\ (2003) used an even  higher density of
4.95$\times{10^{8}}$ cm$^{-3}$ in their model of polarized OH maser emission in W75N. The magnetic field of the maser spots is the general
interstellar field enhanced by gas compression. Both direct and reverse MHD shocks may be generated by recurrent
outflows leading to the density enhancement at maser spot locations. In such a model one is not expecting to observe a large physical motion for the
maser spots in proper motion measurements. Rather, the shock motion causes consecutive brightening of the maser spot objects encountered by the shock.
This is the 'Christmas tree' model of maser emission variations, with lights flashing at fixed positions, as opposed to a model of physical motion for
the maser spots. The maser spot objects can be pre-existing gas condensations which are excited by the passing shock.

\subsubsection{Planets}

 The magnetic field of the maser spots can also be intrinsic to the
maser spot objects. Every maser spot object may have its own magnetic field originating in a dense magnetized, solid or liquid core such as a
rotating planet orbiting the central star. The maser emission is generated in an extended water--methanol gas envelope which is formed by sublimation
from the solid icy cover of the planet. OH molecules are produced from water by dissociation as in comets. Such a model of masers was proposed
by Slysh et al.\ (1999). These envelopes can be energized by MHD shocks from the stellar wind in the same way as in previous model. The maser emission is
generated in those planetary envelopes which have been impacted by the shock at a particular moment.
In this model the masers are identified as icy planets rotating around young massive stars in the proto-planetary disk. The disk is shielded from
ionizing radiation of the star by the ultra-compact
H~II region which rests inside the disk and absorbs all UV-photons, and by a cold dense dust core which absorbs visible and near-IR photons. The core
reemits all the absorbed energy in the far-infrared. This emission may serve as a pump for the masers.
The masers in star forming regions are found to be
associated with protostars or young massive stars
of ZAMS types O and B. Although more than 200 extrasolar planets were discovered
around solar-type stars, little is known about existence of planets orbiting young massive stars. Planets around a pulsar perhaps may serve as an indirect
evidence of planets  belonging to more massive stars (Wolszczan \& Frail 1992). Dusty disks with possible crystalline grains were discovered around two very massive stars in LMC
(Kastner et al.\ 2006), at the distance between 120 and 2500~au from the stars. The Kuiper Belt-like structures composed of debris may exist around
these stars, and formation of larger objects -- planets -- can not be excluded.

\section{Conclusions}

 A very strong magnetic field of 40~mG has been detected in several OH maser spots which have appeared during a flare of OH maser emission
 in 2000, within 110~au from the ultra-compact H~II region. The magnetic field probably
 originates in the exciting star where its intensity is about 500~G, or from the compression of interstellar gas by MHD shock, or in icy planetary bodies
 serving as nuclei for the maser spot emission. More frequent high angular resolution observations of future flares may help to distinguish between these models.

\section*{Acknowledgments}

We acknowledge NRAO and EVN for providing efficient access to VLBI archive data. NRAO is a facility of the NSF operated under cooperative agreement
by Associated Universities, Inc. The European VLBI Network is a joint facility of European, Chinese, South African and other radio astronomy institutes
 funded by their national research councils. The work of VIS was supported by RFBR Grant No 04-02-17057a.


\begin{thebibliography}{99}
\bibitem{b1} Alakoz A.V., Slysh V.I., Popov M.V., Val'tts I.E., 2005,  Astron. Letters,  31,  375
\bibitem{b2} Baart E.E., Cohen R.J., Davies R.D., Norris R.P., Rowland P.R., 1986,  MNRAS,  219, 145
\bibitem{b3} Fiebig D., G\"{u}sten R., 1989, A\&A, 214, 333
\bibitem{b4} Fish V.L., Reid M.J., Argon A.L., Zheng X.-W., 2005, ApJS, 160, 220
\bibitem{b5} Gray M.D., Hutawarakorn B., Cohen J.R., 2003, MNRAS, 343, 1067
\bibitem{b6} Haschick A.D., Reid M.J., Burke B.F., Moran J.M., Miller G., 1981, ApJ, 244, 76
\bibitem{b7} Hunter T.R., Taylor G.B., Felli M., Tofani G., 1994, A\&A, 284, 215
\bibitem{b8} Hutawarakorn B., Cohen R.J., Brebner G.C., 2002, MNRAS, 330, 349
\bibitem{b9} Kastner J.H., Buchanan C.L., Sargent B., Forrest W.J., 2006, ApJ, 638, L29
\bibitem{10} Minier V., Conway J.E., Booth R.S., 2001, A\&A, 369, 278
\bibitem{b11} Sarma A.P., Troland T.H., Crutcher R.M., Roberts D.A., 2002, ApJ, 580, 928
\bibitem{b12} Shepherd D.S., Kurtz S.E., Testi L., 2004,  ApJ,  601,  952
\bibitem{b13} Shu F., Najita J., Ostriker E., Wilkin F., Ruden S., Lizano S., 1994, ApJ, 429, 781
\bibitem{b14} Slysh V.I., Migenes V., Val'tts I.E., Lyubchenko S.Yu., Horiuchi S., Altunin V.I., Fomalont E.B., Inoue M., 2002,  ApJ,  564,  317
\bibitem{b15} Slysh V.I., Val'tts I.E., Kalensky S.V., Larionov G.M., 1999, Astron. Reports, 43, 657
\bibitem{b16} Torrelles J.M., G\'{o}mez J.F., Rodr\'{i}guez L.F., Ho P.T.P., Curiel S., V\'{a}zquez R., 1997, ApJ, 489, 744
\bibitem{b17} Torrelles J.M. et al., 2003, ApJ, 598, L115
\bibitem{b18} Uchida Y., Shibata K., 1985, PASJ, 37, 515
\bibitem{b19} Vlemmings W.H.T., Diamond P.J., van Langevelde H.J., Torrelles J.M., 2006, A\&A, 448, 597
\bibitem{b20} Wade G.A., Fullerton A.W., Donati J.-F., Landstreet J.D., Petit P., Strasser S., 2006, A\&A (in press, astro-ph/0601623)
\bibitem{b21} Wolszczan A., Frail D.A., 1992, Nat, 335, 145
\bibitem{b22} Yngvesson K.S., Cardiasmenos A.G., Shanley J.F., Rydbeck O.E.H., Elld\'{e}r J., 1975, ApJ, 195, 91



\end{thebibliography}
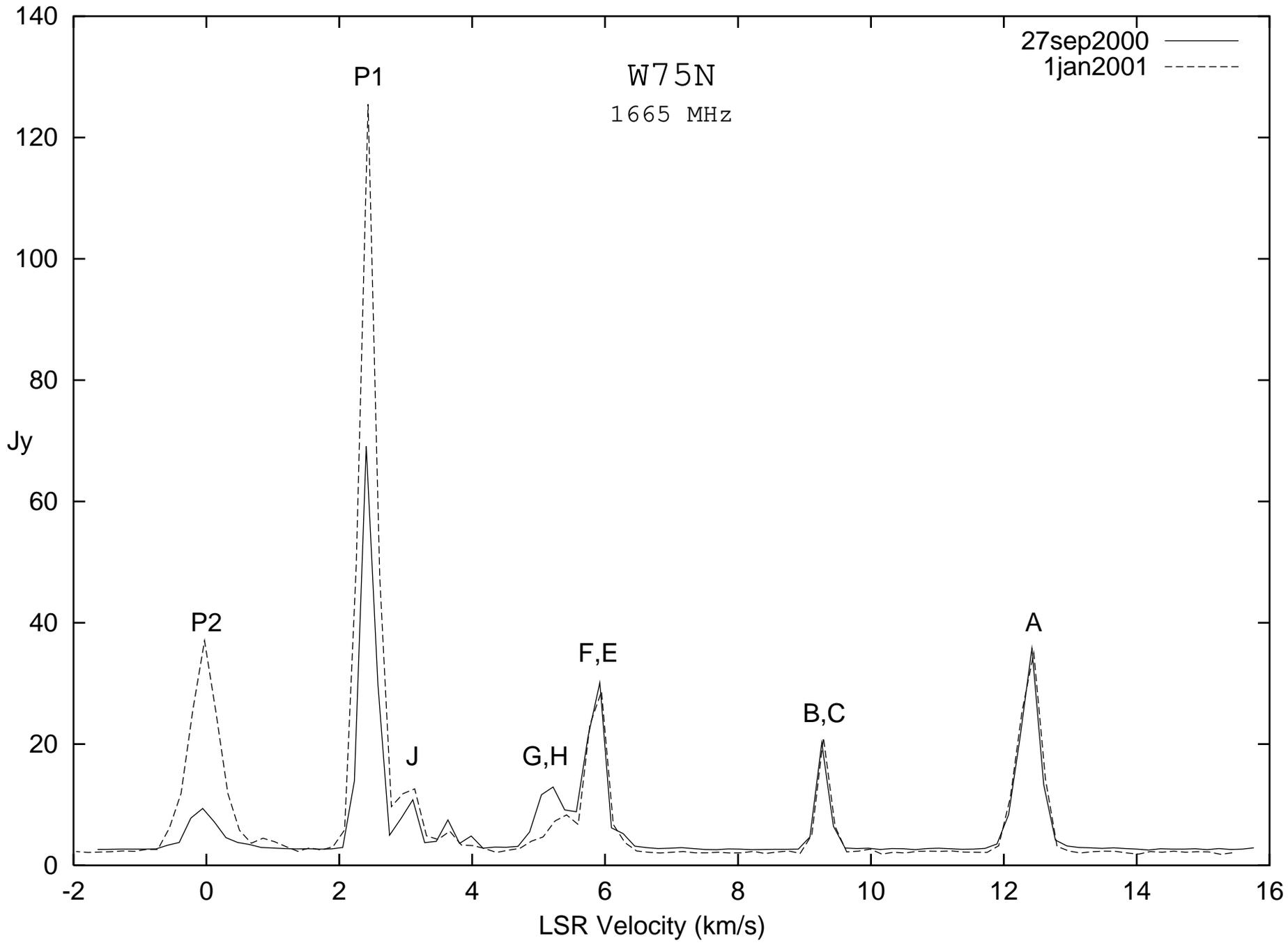
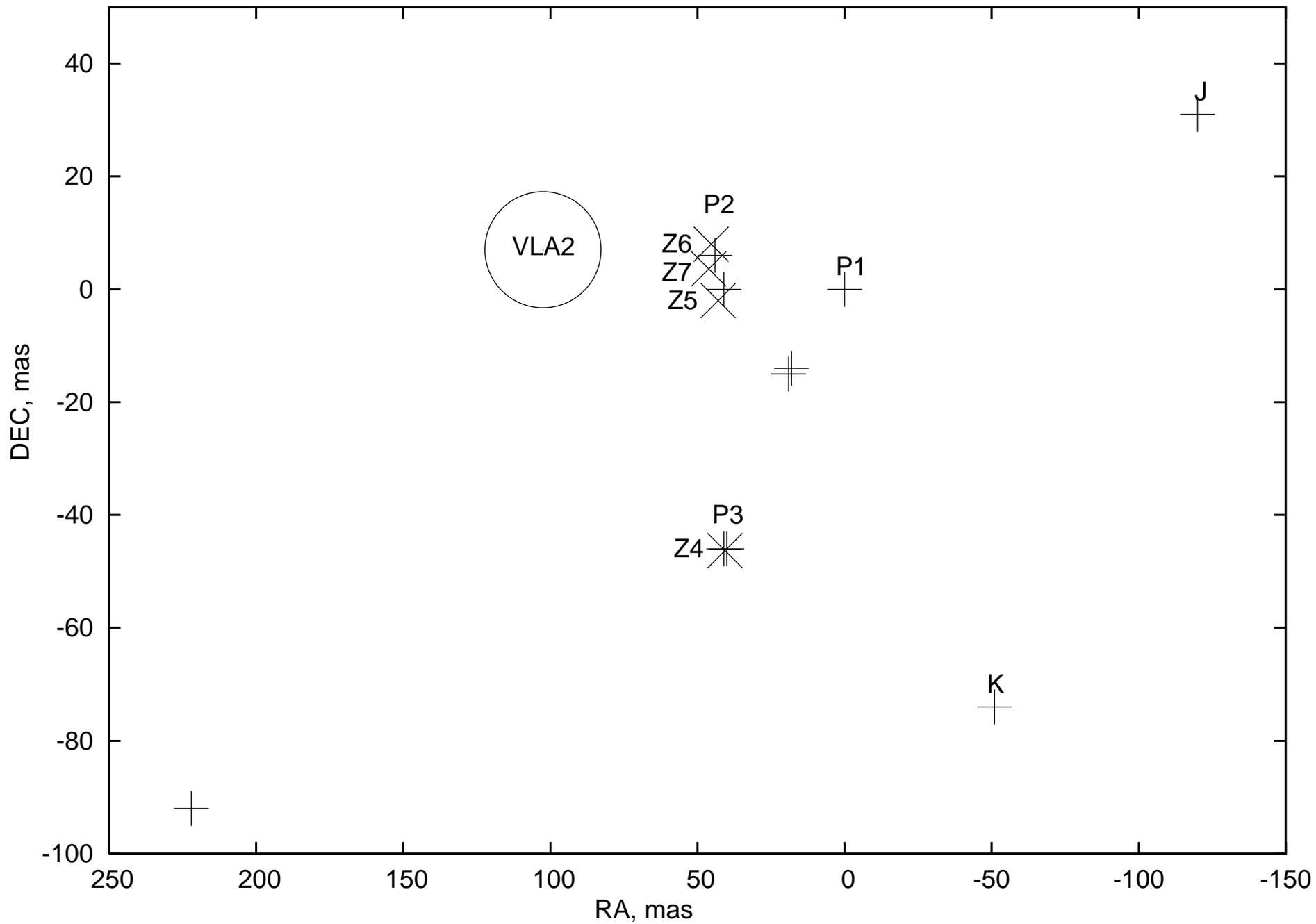
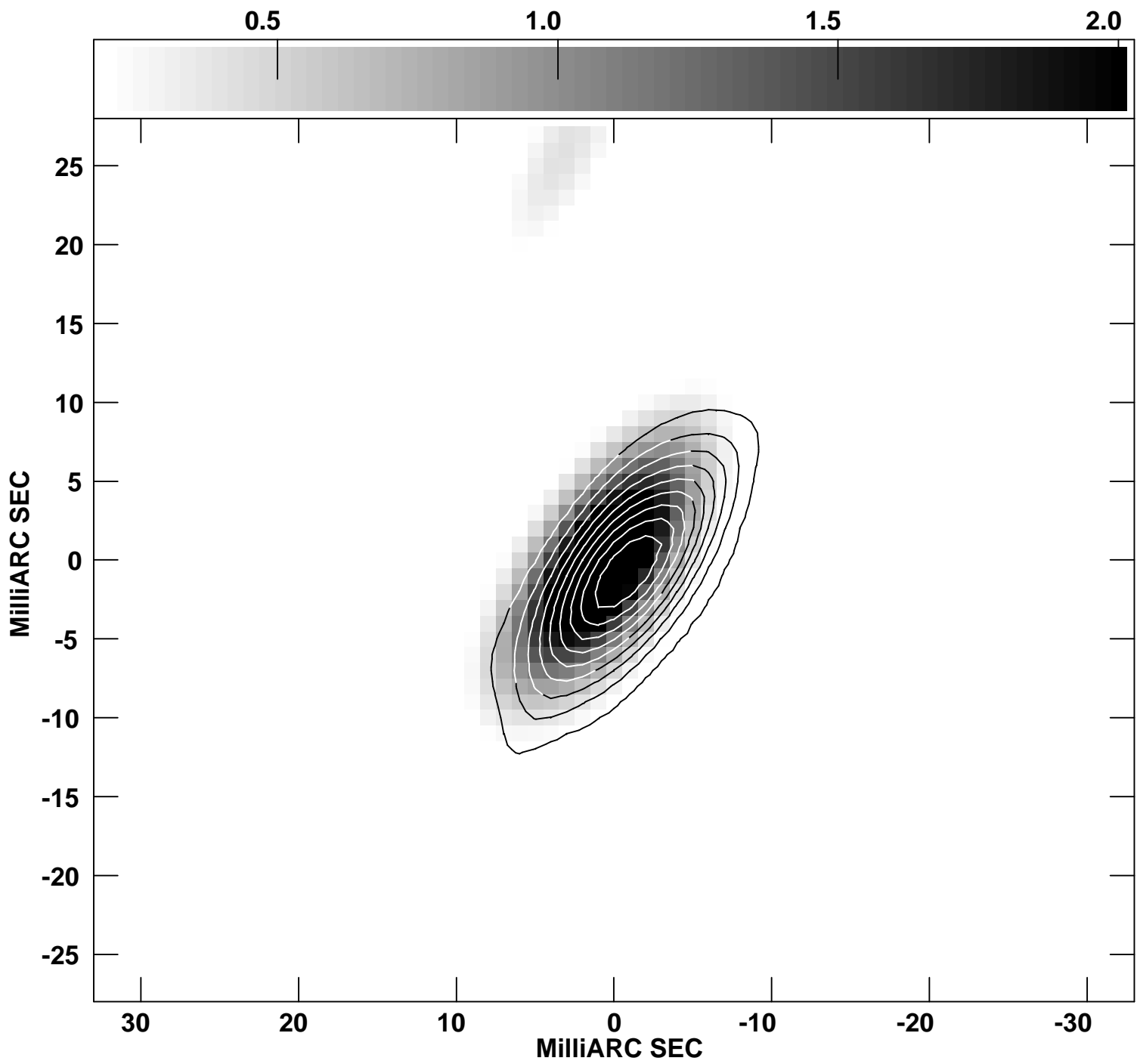
\end{document}